# UTILIZING IMBALANCED DATA AND CLASSIFICATION COST MATRIX TO PREDICT MOVIE PREFERENCES

Haifeng Wang[1]

[1]Penn State University New Kensington, New Kensington, PA, USA.

*ABSTRACT*

*In this paper, we propose a movie genre recommendation system based on imbalanced survey data and unequal classification costs for small and medium-sized enterprises (SMEs) who need a data-based and analytical approach to stock favored movies and target marketing to young people. The dataset maintains a detailed personal profile as predictors including demographic, behavioral and preferences information for each user as well as imbalanced genre preferences. These predictors do not include movies' information such as actors or directors. The paper applies Gentle boost, Adaboost and Bagged tree ensembles as well as SVM machine learning algorithms to learn classification from one thousand observations and predict movie genre preferences with adjusted classification costs. The proposed recommendation system also selects important predictors to avoid overfitting and to shorten training time. This paper compares the test error among the above-mentioned algorithms that are used to recommend different movie genres. The prediction power is also indicated in a comparison of precision and recall with other state-of-the-art recommendation systems. The proposed movie genre recommendation system solves problems such as small dataset, imbalanced response, and unequal classification costs.*

## 1. INTRODUCTION

Data analytics is needed to determine which targeted customer segments are most likely to buy or rent specific genres of movies from physical movie retail chains. Many small-medium entities (SMEs) need a simple analytical framework at hand to determine whom they should be marketing. The goal of this paper is to propose a more useful and practical predictive model to SMEs in dire need of a data-based, analytical approach to finding customers who are more likely to be interested in specific movie genres and subsequently determine which market segments these companies should target in their marketing campaigns based on the predictions.

SMEs will use the predictive analysis to find the right types of new customers as well as help purchase popular movies to stock in their store. An accurate predictive analysis cannot only retain more customers, but it can also reduce the amount of inventory as inventory capitals for long periods of time. The more inventory turnover and customer base are, the better the business is going. The movies are leaving the store quicker, to be bought or rented then the inventory turnover becomes higher and more revenue will be brought in to movie rental companies.

Unlike those predictive models of Redbox and Netflix, this paper does not answer questions regarding the current mega-trends in which demographics and personal attributes contribute to an individual liking certain movie genres across the whole country. But our model is designed to be able to predict genre preference within a city or a state. As [1] said, recommender users that live in South America often dislike Hollywood drama movies. Hence a movie audience in different areas have different movie genre preferences and recommendations for an audience in other parts of the world have to be refined. The reason why we did not use 10 million datasets from Movie Lens [2] is that those dataset does not include audiences' detailed demographic information. This information makes our research unique and more customer-oriented.

The Movie Lens dataset only represents user with an ID and does not contain other personal information about their background, hoppy, habit, family and social network [1]. But the user profile is found to be critical to improving prediction accuracy in recommendation systems[3]. Although Group Lens database (https://grouplens.org/datasets/) includes users' gender, age, zip-code and occupation, the young people survey used in this paper includes more details such as hobbies and interests, phobias, health habits, personality traits, spending habits and demographics information of survey participants aged between 15 and 30.

But experiments from physical movie retail chains indicate that imbalanced class distribution makes it difficult to use these classifier learning algorithms that are designed for balanced class distribution [4-6]. The survey about specific movie genres bought or rented by customers has many imbalanced genre distributions. For example, science fiction movies have more instances of favor class than dislike. The favor class is represented by a large of samples while the dislike class is shown by only a few. Standard classifiers such as support vector machines and decision tree do not work well on imbalanced data sets because they are supposed to learn from balanced training data and output the best prediction that fits the data. The hypothesized prediction does not focus on rare cases in an imbalanced dataset. Therefore the few class is predicted to be less and weaker than the other prevalent class. As a result, the few class samples are misclassified in more situations than those prevalent class test samples. Moreover, a more valuable classification model is able to correctly classify rare class samples. For instance, as far as western movies, the favor cases are usually much less than dislike cases. A favorable classification model is one that has a higher identification rate for those western movie lovers. The reported imbalance problem also exists with many other applications [5-7].

Many researchers have been attracted to solve the class imbalance problems in machine learning and data mining fields. The paper [8] reviews three aspects of the imbalance problems: the nature of the problem; possible solutions and performance evaluation in the presence of the class imbalance problem. Some papers proposed to resample data space to rebalance the class distribution by oversampling instances of the small class and under-sampling prevalent class disadvantages. But the disadvantages are obvious because the under-sampling may lose information on the prevalent class and the resampling may not get the optimal class distribution of a training data space. Although it is hard to state explicitly what imbalance degree can deteriorate the classification performance, in some applications a ration as low as 1:10 is hard to build a good model[8]. Therefore more researches focus on the classifier learning algorithm to advance the classification of imbalanced data.

The objective of this paper is to investigate the classification of imbalanced data because the prediction model in the previous article [9] is not always satisfactory when the model is used to deal with unbalanced sample data for prediction of customers' movie genre preferences. Although SVM does well with imbalanced data to some extent, SVM is very limited when it is used to learn from imbalanced datasets in which negative instances heavily outnumber the positive instances[10]. The adapted ensemble learning algorithms such as AdaBoost and bagged trees have better performance by introducing cost items into the learning framework. The cost items represent uneven identification importance between classes and are passed as a square matrix with nonnegative elements to the algorithms.

This paper has three advantages as below:

1. This paper finds and explores the features that are important for movie genre preference prediction.
2. The proposed method effectively handles the problems caused by a small dataset, imbalanced response data, and unequal classification costs.
3. The proposed method is competitive against other state-of-the-art recommendation systems in test accuracy.

The paper is structured as follows. In section 2, we review related work. In section 3, we describe the machine learning algorithms used in our experiments. In section 4, we evaluate and discuss experimental results and comparison with other state-of-art predictive models. In section 5, we leave the reader with concluding thoughts and future work.

## 2. RELATED WORK

Recommendation performance has been improved recently by different approaches [11-14]. Researchers usually categorize recommender systems into collaborative filtering [15] and content-based filtering systems [11, 16, 17]. A brief review of both filtering methods and known issues associated with the approaches have been summarized in [9]. The proposed work here can be viewed as a continuation of the works in [9]. We summarized some key points from existing customer-oriented recommender systems and presented new contributions. In particular, we extended the previous application by helping SMEs utilize data analysis to answer questions regarding customer preference and marketing segment of the movie rental industry.

## 3. METHODS

In this section, we describe the methodology and discuss the prediction power of the proposed recommendation system.

### 3.1 Handle data

The processed dataset consists of one sample with 1007 observations regarding movie genres. It also includes preferences for 11 types of movies, 22 personal attribute variables, and 5 demographic variables. Using these observations, we will display the relationships between movie preferences and demographic information including age, gender and other factors that can be used to accurately recommend specific genres to customers as other recommendation systems did [18-21]. It has been approved that user information such as gender, location, or preference is effective to recommend movies to the customer [3, 22-24]. In this paper, we will examine more characteristics of survey respondents for different movie genres because the dataset of genres is imbalanced.

The dataset of 1007 respondents as shown in [9] with quota characteristics enabled the research to be generalized to the young population. The sample was composed of 38 hypothetical decision-making dimensions shown in Table 1 (11 for different preference to movie genres; 27 for demographic information) with a rating ordinal scale from 1 to 5. These predictors allowed us to perform analysis and prediction based on both qualitative and quantitative aspects of the domain. The scale was set as below: 1-hate, 2-dislike, 3-acceptable, 4-like, 5-love. For the sake of simplicity, we only regarded the scale of 4 and 5 as a positive recommendation and other scales were negative ones. The selection of a full set of features in the dataset is explained in [9]. And we do not train the proposed model with other information such as actor information provided in Group lens dataset because such features have been fully explored in existing recommendation systems.

Table1. Hypothetical decision-making dimensions for classification in machine learning

| Movies number | Fantasy | Shopping |
|---|---|---|
| Horror | Animated | Children |
| Thriller | Documentary | Gender |
| Comedy | Theatre | Reading |
| Romantic | Friends Number | Struggles |
| Countryside | Western | Education |
| Sci-fi | Action | Happiness |
| War | Socializing | Child Number |
| Outdoors | Energy | Village |
| Only Child | Loneliness | Hobbies |
| Leisure activity | Online Chat | Age |
| Finances | Fun with Friends | Internet usage |
| Entertainment Cost | PC usage | |

### 3.2 Bootstrap aggregation Ensemble classifier

Ensemble classifier is a machine learning algorithm. It is used for classification in this paper because it is more resistant to overfitting than the above classifiers. It is also called multi-classifier since it uses multiple iterations to create a strong classifier by adding weak learners iteratively. In each of the iterations of training, the ensemble will add a new weak learner and use a weighting vector focus on observations that were misclassified in the previous round. So the final classifier has higher accuracy than weak classifiers. And it decreases variance, especially in the case of unstable classifiers, and may produce a more reliable classification than the above single classifiers.

Bagging, which stands for "bootstrap aggregation" and Boosting are both Multi-classifier methods. This paper chooses a classification tree as a base learner algorithm. Multi-classifier gets N learners by generating additional data in the training stage. Bagging algorithm samples training data randomly from the original set to generate additional data and generate multiple versions of a predictor and use these. Bagging method trains each model in parallel because they are independent with each other and then get an aggregated predictor. When the aggregated predictor predicts a numerical outcome, it averages over the different versions and does a plurality vote to predict a class. Figure.1 shows how the ensemble error changes with the accumulation of trees.

The ensemble classifier is different from SVM because it only chooses those features known to improve the predictive power of the learner. The bagged decision trees select observations with replacement with omits on average 37% of data for each decision tree. These are called "out-of-bag". The algorithm compares the out-of-bag prediction response against the observed responses for all training data to estimate the average out-of-bag error. By randomly permuting out-of-bag observations across one feature at a time and estimating, there is an increase in the out-of-bag error due to this permutation. If the increase is larger, the feature should be more important. It is a very attractive feature of the bagged tree because it can reduce the feature dimensionality and potentially shorten training time as irrelevant features are disregarded.

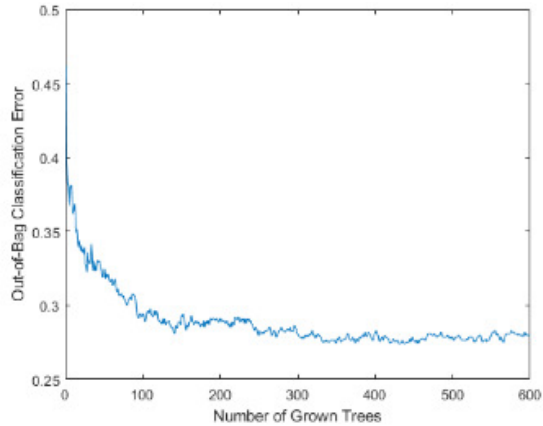

Fig.1 Out-of-Bag (OOB) Classification error decreases to about 0.27 after with 400 grown trees

The proposed approach uses bagged decision trees for classification and estimate feature importance shown in Figure.2. This paper selects the features yielding an importance measure greater than 0.1. And this threshold is chosen arbitrarily. The selected features include:{"Online chat", "Outdoors", "Leisure activity", "Computer usage", "Shopping", "Number of Friends", "Socializing", "Internet usage", "Entertainment spending", "gender", "Only Child"}. Having selected the most important features, a larger ensemble grows on the reduced feature set with less error and the new OOB error is shown in Figure. 3.

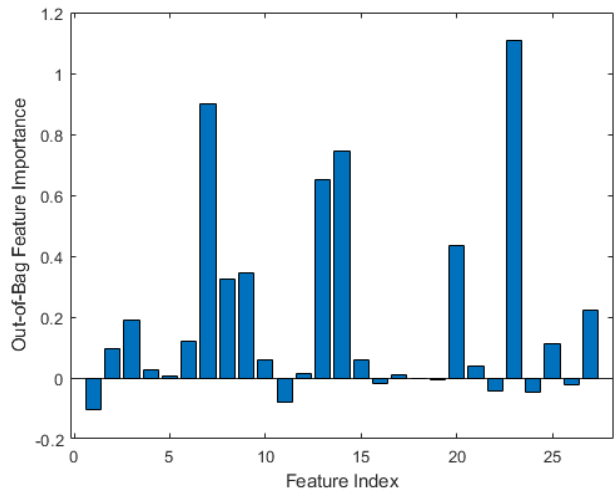

Fig.2 Estimate and selected the most important features using bootstrap aggregation of classification trees

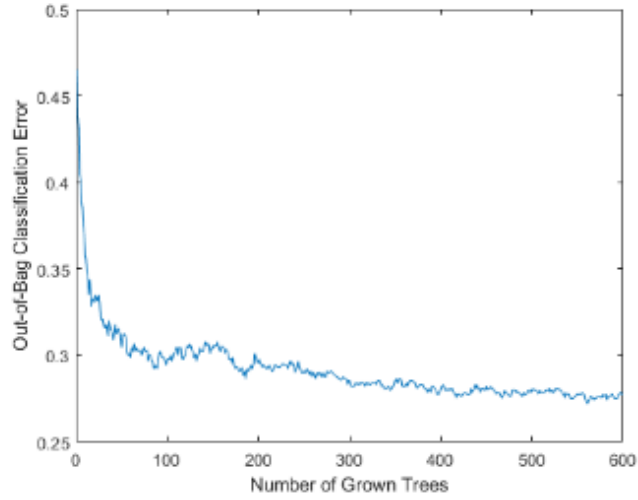

Fig.3 New OOB error generated by new ensemble grows on the reduced feature set

### 3.3 AdaBoost algorithm derivation

AdaBoost is an adaptive boosting ensemble classification learning algorithm that iterates multiple times to generate one composite binary learner in this application. The total error of a boosted classifier $C_{m-1}(x_i)$ after m-1 iteration is the sum of its exponential loss on each data point $x_i$ as follows:

$$E = \sum_{i=1}^{N} \omega_i^{(m)} e^{-y_i \alpha_m k_m(x_i)} \quad (1)$$

Where $\omega_i^{(1)} = 1$ and $\omega_i^{(m)} = e^{-y_i \alpha_m C_{m-1}(x_i)}, \text{ for } m > 1$

**Let's choose the classifier $k_m$ in each iteration that minimized the total weighted error**

$$\sum_{y_i \neq k_m(x_i)} \omega_i^{(m)}. \quad (2)$$

The error rate is defined as:

$$\epsilon_m = \frac{\sum_{y_i \neq k_m(x_i)} \omega_i^{(m)}}{\sum_{i=1}^{N} \omega_i^{(m)}} \quad (3)$$

The weight:

$$\alpha_m = \frac{1}{2} \ln\left(\frac{1-\epsilon_m}{\epsilon_m}\right). \quad (4)$$

The improved boosted classifier:

$$C_m = C_{m-1} + \alpha_m k_m. \quad (5)$$

## 3.4 Which is the best, Bagging or Boosting?

Bagging and Boosting use sampling-learning-combination methods, but there are some differences in details. For example, each training set in Bagging is irrelevant. That is, each base classifier is irrelevant. The training set in Boosting needs to be adjusted in the previous round, which also makes it impossible to calculate in parallel. The prediction function in Bagging is uniform and equal, but in Boosting the prediction function is weighted.

Compared with Bagged tree ensemble, Boosting generates a combined model with lower errors as it optimizes the advantages and reduces pitfalls of the single model. Although Boosting algorithm doesn't help to avoid over-fitting, we have fixed over-fitting by adjusting VC-dimensions in a previous paper and we use early stopping strategy to reduce overfitting, therefore, AdaBoost is the best option. To avoid overtraining, we use five-fold cross-validated ensembles in the paper.

## 3.5 Train Ensemble with Unequal Classification Costs

In this movie genre classification application, it is preferred to treat recommendation classes asymmetrically. For example, there are more people who like comedy movies than others who do not like. And there are more people dislike western movies than those who like. The following descriptive statistics table 2 lists imbalanced movie genre data in a customer survey.

Table 2. Movie genre preference distribution

| Movie Genre | Class value | Count | Percentage | Classified class | imbalance degree (Ratio) |
|---|---|---|---|---|---|
| Science fiction | 1 | 3 | 0.34% | Dislike | 1:10.5 |
|  | 2 | 9 | 1.01% |  |  |
|  | 3 | 65 | 7.30% |  |  |
|  | 4 | 183 | 20.56% | Favor |  |
|  | 5 | 630 | 70.79% |  |  |
| Comedy | 1 | 2 | 0.22% | Dislike | 1:8 |
|  | 2 | 20 | 2.25% |  |  |
|  | 3 | 77 | 8.65% |  |  |
|  | 4 | 220 | 24.72% | Favor |  |
|  | 5 | 571 | 64.16% |  |  |
| Western | 1 | 329 | 36.97% | Dislike | 7:1 |
|  | 2 | 279 | 31.35% |  |  |
|  | 3 | 169 | 18.99% |  |  |
|  | 4 | 68 | 7.64% | Favor |  |
|  | 5 | 45 | 5.06% |  |  |

Based on the above table, the western movie genre has fewer observations of "like" class and more observations of "dislike" class while science and comedy movies have fewer observations of "dislike" class and more observations of "like" class. In the real world, the two classes of the three genres are mixed in the same proportion. So we do not need set prior probabilities for "like" class and "dislike" classes. The three genres are adequately represented by data but do not have balanced observation in positive and negative classes.

Besides attention to positive prediction error, the negative prediction error is also important. Misclassifying observation of "like" has more severe economic consequences than misclassifying observations of "dislike". So we want to treat them asymmetrically by cost parameter. Suppose you want to classify "like comedy movie" and "dislike comedy movie" in an audience. Failure to identify a customer (false negative) has more consequences than misidentifying "like comedy movie" customer as "dislike comedy movie" customer (false positive) since shops will waste time and money on more comedy movies than needed, which cannot bring more customers who like comedy movie and loss other customers. So high-cost should be assigned to misidentifying "dislike comedy movie" as "like comedy movie" and low cost to misidentifying "like comedy movie" as "dislike comedy movie".

The algorithm passes misclassification costs as a square matrix with nonnegative elements. Element C (m, n) of the matrix is the cost of classifying an observation into class j if the true class is i. The diagonal elements C (m, n) of the cost matrix must be 0. For choosing "dislike comedy movie" to be class 0 and "like comedy movie" to be class 1. Then the cost matrix is set to be $\begin{bmatrix} 0 & t \\ 1 & 0 \end{bmatrix}$. Where $t > 1$ is the cost of misidentifying "dislike comedy movie" to "like comedy movie". Since the costs are relative, multiplying all costs by the same positive number does not affect the result of classification. In this application, there are only two classes. So multiplying the cost matrix is equivalent to adjusting their prior probabilities.

GentleBoost is also called as Gentle AdaBoost because it combines features of AdaBoostM1 and Logit Boost. It is a good candidate for binary classification of data. GentleBoost can also classify data with many categorical predictor levels and binary responses.

Table 3. Confusion matrix with different classification costs

| Algorithms | GentleBoost Algorithm | AdaBoost-M1 Algorithm |
|---|---|---|
| Classification Cost | Confusion Matrix for Comedy movie | Confusion Matrix for Comedy movie |
| $\begin{bmatrix} 0 & 1 \\ 1 & 0 \end{bmatrix}$ | $\begin{bmatrix} 105 & 169 \\ 109 & 409 \end{bmatrix}$ | $\begin{bmatrix} 104 & 170 \\ 81 & 437 \end{bmatrix}$ |
| $\begin{bmatrix} 0 & 5 \\ 1 & 0 \end{bmatrix}$ | $\begin{bmatrix} 222 & 52 \\ 332 & 186 \end{bmatrix}$ | $\begin{bmatrix} 219 & 55 \\ 169 & 349 \end{bmatrix}$ |
| $\begin{bmatrix} 0 & 2 \\ 1 & 0 \end{bmatrix}$ | $\begin{bmatrix} 149 & 125 \\ 203 & 315 \end{bmatrix}$ | $\begin{bmatrix} 147 & 127 \\ 169 & 349 \end{bmatrix}$ |

We use the two types Adaboost ensemble algorithms to predict customer preference as comedy movie. Without using a classification cost matrix that reflects this belief, the AdaBoost-M1ensemble predict better than the GentleBoost Algorithm as shown in the third row in Table 3.

In 792 observations, 518 customers like the comedy genre. The AdaBoost-M1 ensemble predicts correctly that 437 customers will like. But for the 274 customers who dislike, the ensemble only predicts correctly that about 60% of customers will dislike comedies. The two types of error in the predictions of the ensemble are defined as below:

- Predicting that the customer likes, but the customer dislikes
- Predicting that the customer dislikes, but the customer likes

International Journal of Artificial Intelligence and Applications (IJAIA), Vol.9, No.6, November 2018

The first error is estimated to be five times worse than the second, so the classification cost matrix is created as $\begin{bmatrix} 0 & 5 \\ 1 & 0 \end{bmatrix}$ as a result of trials. The misclassification cost is then used in the AdaBoost-M1 ensemble to generate the confusion matrix at the fourth row in Table 3. As expected, the new predictor does a better job to classify the customer who dislikes comedy movies. Based on the Trial and error method, the paper also compared confusion matrixes generated by different cost matrixes to find the best cost matrix for the move genre prediction.

## 4. EVALUATION

### 4.1 Test results and discussion

We trained the classifiers with the classification cost matrix $\begin{bmatrix} 0 & 5 \\ 1 & 0 \end{bmatrix}$ to predict the preference at three movie genres. The test results are shown in Table 4, 5 and 6,

Table 4. Science movie prediction error in training and tests

| Error | Machine learning Algorithm For Science movie | | |
|---|---|---|---|
| | *AdaBoost* | *Bagged Tree* | *SVM* |
| Error-out-sample | 0.162 | 0.181 | 0.355 |
| Error-in-sample | 0.17 | 0.194 | 0.286 |

Table 5. Western movie prediction error in training and tests

| Error | Machine learning Algorithm For Western movie | | |
|---|---|---|---|
| | *AdaBoost* | *Bagged Tree* | *SVM* |
| Error-out-sample | 0.125 | 0.13 | 0.143 |
| Error-in-sample | 0.14 | 0.14 | 0.19 |

Table 6. Comedy movie prediction error in training and tests

| Error | Machine learning Algorithm For Comedy movie | | |
|---|---|---|---|
| | *AdaBoost* | *Bagged Tree* | *SVM* |
| Error-out-sample | 0.285 | 0.30 | 0.31 |
| Error-in-sample | 0.262 | 0.285 | 0.29 |

In order to evaluate different aspects of classifiers, we have also used Information Retrieval Performance metrics. Certainly, different applications have different precedence to precision and recall. In our application, a recall was frequently regarded as more important than precision, as it was acceptable to increase the number of false positives (FP). Some customers may also like to rent or buy comedy movies for entertainment even though their responses were a scale of 3

International Journal of Artificial Intelligence and Applications (IJAIA), Vol.9, No.6, November 2018(acceptable) regarded as a negative prediction in our algorithm. The test summary is listed in Table 7.

Table 7. A summary of test performance.

| Genre type | Precision | Recall | Accuracy |
|---|---|---|---|
| Science movie | 0.93 | 0.95 | 0.84 |
| Western movie | 0.92 | 0.88 | 0.88 |
| Comedy movie | 0.67 | 0.86 | 0.72 |

It is preferable to find more customers among those who prefer to a movie genre such as the comedy and provide favored movies of that genre to them. So the prediction performance of our predictive models enables SMEs to answer questions regarding which customers are more likely to be interested in specific movie genres and subsequently as well as which market segments ought these companies target in their marketing campaigns based on the data. In sum, we focused the analysis of the performance by prioritizing recall over precision and we focused the analysis on positive recommendations (in much favor of certain genre) rather than on negative ones.

### 4.2 Comparison with other state-of-art methods used in movie recommendation

We provided a comparison between our work and others to complete the analysis of our proposal. We have not found any example in the literature using exactly the same dataset employed in this work to recommend movie genres to SMEs. But we have noticed that Grouplens has also developed a dataset with limited user profiles similar to ours. Grouplens dataset includes users' gender, age, zip-code, and occupation. Although the features in the dataset we used are more than 5 times larger than the one provided by Grouplens, we provide an approximate comparison with All-Postulates methods (AP) and a hybrid approach based on Estimation of Distribution Algorithms [1, 25]. The test set used for the comparison was obtained from the Grouplens Dataset and included 3000 observations. To ensure that unrelated features being included in the dataset had no influence on the prediction made, we removed the corresponding record from the dataset such as actors' information.

The All-Postulates methods can account for several different aspects and genres considered together through a dialectical analysis. This recommendation system can also stem for either content-based or collaborative filtering techniques or both. The other recommendation system based on Estimation of Distribution Algorithms applies the Estimation of Distribution Algorithms (EDAs) to learn users' preferences and accurately describe users' interest features. Based on the user interest profiles, the model is able to recommend movie genres.

Figure.4 shows a performance comparison of the above methods. In this figure, we report the best performance calculated by each of the above approaches in terms of precision and recall. From the comparison, we can conclude that the effectiveness of our machine learning model (SME) is not inferior to that of the other two models although the small test dataset does not have the same predictors as what we used in the research.

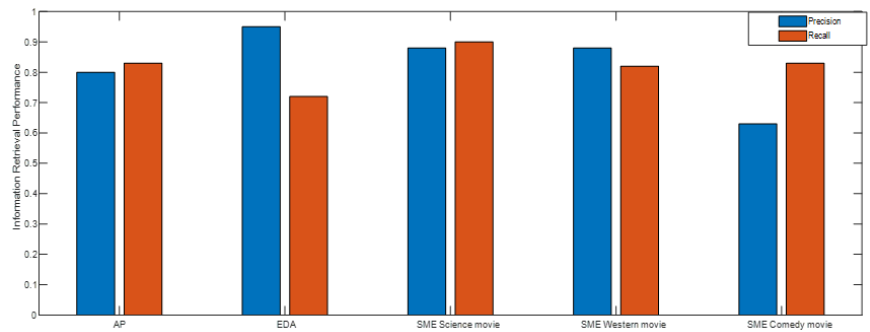

Fig.4. A performance comparison chart with other state-of-the-art methods on movie recommendation systems. AP represents argument-based mixed recommenders; EDA refers to a hybrid approach based on Estimation of Distribution Algorithms; SME is our recommendation engine for small and Medium-sized Enterprises.

## 5. CONCLUSION

Large corporations such as Netflix and Redbox have gained a great competitive edge from collecting and analyzing customer demographic information to find customer interest and target market segments. Many small-medium entities (SMEs) in physical movie rental industry also need localized data-based analytical and recommendation systems to determine which targeted customer segments are most likely to buy or rent specific genres of movies. From our studies that were performed, we can conclude that the application of machine learning techniques for movie genre prediction is quite successful with a small dataset. The proposed predictive model analysed specific demographic information and users profile information to make prediction of movie genre preference for SMEs in the movie rental industry. The experiments showed that the adjusted bagged decision trees ensemble model can be used to accurately predict local customers' movie genre preference such as scientific, comedy and western movies based on a small dataset and classification cost. Finally, in comparison with other state-of-art prediction models, the proposed prediction model has satisfied prediction power. Therefore the proposed machine learning algorithm allows us to overcome shortcomings of traditional massive dataset prerequisite and can be more flexible applied based on the misclassification cost. We will continue to improve the algorithm and enrich the dataset because the small dataset is not enough to predict the current mega-trends in which demographics and personal attributes contribute to individual movie genre preference across the country.

## REFERENCES

[1] C. E. Briguez, M. C. D. Budán, C. A. D. Deagustini, A. G. Maguitman, M. Capobianco, and G. R. Simari, "Argument-based mixed recommenders and their application to movie suggestion," Expert Systems with Applications, vol. 41, no. 14, pp. 6467-6482, 2014/10/15/ 2014.

[2] F. M. Harper and J. A. Konstan, "The MovieLens Datasets: History and Context," ACM Trans. Interact. Intell. Syst., vol. 5, no. 4, pp. 1-19, 2015.

[3] S.-M. Choi, S.-K. Ko, and Y.-S. Han, "A movie recommendation algorithm based on genre correlations," Expert Systems with Applications, vol. 39, no. 9, pp. 8079-8085, 2012/07/01/ 2012.

[4] N. V. Chawla, N. Japkowicz, and A. Kotcz, "Editorial: special issue on learning from imbalanced data sets," SIGKDD Explor. Newsl., vol. 6, no. 1, pp. 1-6, 2004.

[5] T. Fawcett and F. Provost, "Adaptive Fraud Detection," Data Mining and Knowledge Discovery, journal article vol. 1, no. 3, pp. 291-316, September 01 1997.

[6] M. Kubat, R. C. Holte, and S. Matwin, "Machine Learning for the Detection of Oil Spills in Satellite Radar Images," Machine Learning, journal article vol. 30, no. 2, pp. 195-215, February 01 1998.

International Journal of Artificial Intelligence and Applications (IJAIA), Vol.9, No.6, November 2018[7]   P. Riddle, R. Segal, and O. Etzioni, "REPRESENTATION DESIGN AND BRUTE-FORCE INDUCTION IN A BOEING MANUFACTURING DOMAIN," Applied Artificial Intelligence, vol. 8, no. 1, pp. 125-147, 1994/01/01 1994.
[8]   Yanminsun, A. Wong, and M. S. Kamel, Classification of imbalanced data: a review. 2011.
[9]   H. Wang and H. Zhang, "Movie genre preference prediction using machine learning for customer-based information," in 2018 IEEE 8th Annual Computing and Communication Workshop and Conference (CCWC), 2018, pp. 110-116.
[10]  V. Ganganwar, "An overview of classification algorithms for imbalanced datasets."
[11]  G. Adomavicius and A. Tuzhilin, "Toward the next generation of recommender systems: a survey of the state-of-the-art and possible extensions," IEEE Transactions on Knowledge and Data Engineering, vol. 17, no. 6, pp. 734-749, 2005.
[12]  D. H. Park, H. K. Kim, I. Y. Choi, and J. K. Kim, "A literature review and classification of recommender systems research," Expert Systems with Applications, vol. 39, no. 11, pp. 10059-10072, 2012/09/01/ 2012.
[13]  A. Said and A. Bellogín, "Coherence and inconsistencies in rating behavior: estimating the magic barrier of recommender systems," User Modeling and User-Adapted Interaction, journal article April 13 2018.
[14]  K. Jasberg and S. Sizov, "The Magic Barrier Revisited: Accessing Natural Limitations of Recommender Assessment," presented at the Proceedings of the Eleventh ACM Conference on Recommender Systems, Como, Italy, 2017.
[15]  A. E. Sarabadani Tafreshi, A. Sarabadani Tafreshi, and A. L. Ralescu, "Ranking Based on Collaborative Feature Weighting Applied to the Recommendation of Research Papers," (in en), International Journal of Artificial Intelligence & Applications, vol. 9, p. 53, 2018.
[16]  P. Lops, M. De Gemmis, and G. Semeraro, "Content-based recommender systems: State of the art and trends," in Recommender systems handbook: Springer, 2011, pp. 73-105.
[17]  J. L. Herlocker, J. A. Konstan, A. Borchers, and J. Riedl, "An algorithmic framework for performing collaborative filtering," in Proceedings of the 22nd annual international ACM SIGIR conference on Research and development in information retrieval, 1999, pp. 230-237: ACM.
[18]  D. Godoy and A. Corbellini, "Folksonomy Based Recommender Systems: A State of the Art Review," International Journal of Intelligent Systems, vol. 31, no. 4, pp. 314-346, 2016.
[19]  P. De Meo, G. Quattrone, and D. Ursino, "A query expansion and user profile enrichment approach to improve the performance of recommender systems operating on a folksonomy," User Modeling and User-Adapted Interaction, journal article vol. 20, no. 1, pp. 41-86, February 01 2010.
[20]  H. Yamaba, M. Tanoue, K. Takatsuka, N. Okazaki, and S. Tomita, "On a Serendipity-oriented Recommender System based on Folksonomy and its Evaluation," Procedia Computer Science, vol. 22, pp. 276-284, 2013/01/01/ 2013.
[21]  H. Liang, Y. Xu, Y. Li, and R. Nayak, "Personalized recommender system based on item taxonomy and folksonomy," presented at the Proceedings of the 19th ACM international conference on Information and knowledge management, Toronto, ON, Canada, 2010.
[22]  R. M. Bell and Y. Koren, "Lessons from the Netflix prize challenge," SIGKDD Explor. Newsl., vol. 9, no. 2, pp. 75-79, 2007.
[23]  D. Billsus and M. J. Pazzani, "Learning Collaborative Information Filters," presented at the Proceedings of the Fifteenth International Conference on Machine Learning, 1998.
[24]  B. Sarwar, G. Karypis, J. Konstan, and J. Riedl, "Analysis of recommendation algorithms for e-commerce," presented at the Proceedings of the 2nd ACM conference on Electronic commerce, Minneapolis, Minnesota, USA, 2000.
[25]  Y. L. T. Liang, J. Fan, J. Zhao, "A hybrid recommendation model based on estimation of distribution algorithms," Journal of Computational Information Systems, vol. 10 (2), pp. 781-788, 2014.
12